**Title:** A bootstrap approach for validating the number of groups identified by latent class growth models

**Authors:** Mésidor M.[1,2], Sirois C.[2,3,4,5], Simard M.[1,4], Talbot D.[1,2]

**Affiliations :**

[1]Département de médecine sociale et préventive, Université Laval, Québec, Canada

[2]Centre de recherche du CHU de Québec – Université Laval Québec, Canada

[3]Faculté de Pharmacie, Université Laval, Québec, Canada

[4]Institut national de santé publique du Québec, Québec, Canada

[5] Centre d'excellence sur le vieillissement de Québec, VITAM – Centre de recherche en santé durable, Québec, Canada

**Corresponding author :** Miceline Mésidor, Département de médecine sociale et préventive, Université Laval, Québec, Canada

E-mail : miceline.mesidor.1@ulaval.ca

**Abstract**

The use of longitudinal finite mixture models such as group-based trajectory modeling has seen a sharp increase during the last decades in the medical literature. However, these methods have been criticized especially because of the data-driven modelling process which involves statistical decision-making. In this paper, we propose an approach that uses bootstrap to sample observations with replacement from the original data to validate the number of groups identified and to quantify the uncertainty in the number of groups. The method allows investigating the statistical validity and the uncertainty of the groups identified in the original data by checking if the same solution is also found across the bootstrap samples. In a simulation study, we examined whether the bootstrap-estimated variability in the number of groups reflected the replication-wise variability. We also compared the replication-wise variability to the Bayesian posterior probability. We evaluated the ability of three commonly used adequacy criteria (average posterior probability, odds of correct classification and relative entropy) to identify uncertainty in the number of groups. Finally, we illustrated the proposed approach using data from the Quebec Integrated Chronic Disease Surveillance System to identify longitudinal medication patterns between 2015 and 2018 in older adults with diabetes.

1. INTRODUCTION

There has been a sharp increase in the use of longitudinal finite mixture models in the last few decades[1]. Because of their ability to characterize heterogeneity over time without the need for predictor variables, these methods have been preferred by many analysts over traditional ones such as mixed models. Different terminologies have been used to refer to the various types of longitudinal finite mixture models. Latent class growth models (LCGMs), also called growth mixture models, and latent class growth analysis (LCGA) are the two of the most common longitudinal finite mixture models[2]. LCGMs are more complex models that feature LCGA as a special case.

Several concerns have been raised amidst the growing use of these methods[3-8]. Studies have shown that the number of latent groups and their shape are influenced by the length of follow-up, with longer durations being associated with more groups[7]. Data spacing between time points and larger variability in the outcome due to measurement error are other factors that have been observed to affect the number of groups found[7, 8]. Finally, when the analysis assumes a normally distributed variable, violation of this assumption has been shown to lead to a greater number of groups than actually exist in the data[4]. Thus, it appears essential to quantify the uncertainty of the number of groups obtained in latent class analyses.

One of the particularities of LCGMs is related to the data-driven modelling process which involves statistical decision-making[9, 10]. The availability of only one sample of data in practice raises concerns about the validity of the number of groups identified, as it is not possible to assess if a different sample from the same population would lead to finding the same number of groups.

The usual tools to assess model adequacy do not allow quantifying this uncertainty. Although statistical procedures are often used to assess the validity of statistical models, few studies have reported the use of validation methods to assess the uncertainty concerning trajectory groups identified in latent class models. In this paper, we propose a novel approach based on bootstrap to: i) validate the number of groups identified from latent class methods and ii) quantify the uncertainty in the identification of the number of groups using simulated data.

Bootstrapping has previously been used as a validation procedure in longitudinal finite mixture models, but, to the best of our knowledge, never to assess the uncertainty in the number of groups. For example, some studies have proposed using the bootstrap to quantify the uncertainty in the classification of observations (individuals) among latent classes or clusters[11]. As another example, Ferro and Speechley used the bootstrap technique to validate the trajectories of depressive symptoms[12]. Because the final dataset analyzed was the average of the 1000 bootstrap samples, they were unable to identify variability between different datasets. Grün and Leisch have also proposed a bootstrap technique, but to analyze the stability of the estimated parameters and to detect identifiability problems[13].

In the following, we first review the notation for LCGMs before introducing our proposed method. A simulation study is then used to evaluate the performance of our approach and compare it with alternatives. We also illustrate the proposed approach in real data to identify longitudinal medication patterns in people aged 75 to 80 years with diabetes.

## 2. METHODS

## 2.1 Latent class growth models

LCGMs are an application of finite mixture modeling, which means that these models include a combination of two or more probability groups. In LCGMs, each group is often described by a polynomial function of time[2]. This paper focuses on the continuous outcome case, but LCGMs allow the specification of other distributions for the outcome such as Poisson or binomial.

For *K* groups, the marginal probability distribution of a randomly chosen group is defined as follows:

$$P(y_i) = \sum_{k=1}^{K} \pi_k P^k(Y_i)$$

where:

$y_i = \{y_{i1}, y_{i2}, \dots, y_{iT}\}$ is the vector of measured continuous outcome for subject $i$, $i = 1, \dots, n$ at time $t$, $t = 1, \dots, T$;

$\pi_k$ is the group membership probability;

$P^k(Y_i)$ is the conditional distribution of the trajectory group, $y_i$, given that the subject *i* is in group *k*.

In the presence of a repeatedly measured continuous outcome, $P^k(Y_i)$ is expressed as a multivariate normal (MVN) density function. For a subject *i* in group *k*,

$$y_i^k \sim MVN(\mu^k, \Sigma^k)$$

where $y_i^k$ is the repeated outcome, $\mu^k$ and $\Sigma^k$ are respectively the mean and covariance matrix in group *k*. It is important to note that $\Sigma^k$ includes both inter-individual and intra-individual variations for LCGM and only intra-individual variations for LCGA.

Although LCGMs allow the outcomes of each subject to be represented as a mixture of the various trajectory groups, it is common in practice to assign each subject to a single group based on the maximum estimated 'posterior' probability. The estimated 'posterior' of subject *i* belonging to class *k* is obtained by applying Bayes' theorem[2, 9]:

$$pp_{ik} = \frac{\hat{\pi}_k \hat{P}^k(y_i)}{\sum_{l=1}^{K} \hat{\pi}_l \hat{P}^l(y_i)}$$

$\hat{P}^k(y_i)$ is the estimated probability of observing the data if the subject *i* is a member of group *k* and $\hat{\pi}_k$ is the estimated proportion of subjects in class *k* with the constraint that $\sum_{k=1}^{K} \hat{\pi}_k = 1$.

One challenge with LCGM is the specification of the number and shape of the groups. There are several statistical fit indices for selecting the number of groups, but the Bayesian Information Criteria (BIC) is the most widely used. A common strategy is to first determine the optimal number of groups by estimating models with increasing number of groups, using the lowest BIC as an indication of the alternative model. Then, the best polynomial order for each group can be identified also based on the BIC[9].

Based on the Guidelines for Reporting on Latent Trajectory Studies (GRoLTS), the adequacy of the model should be evaluated using several criteria[10]. One of them is the average posterior probability (APP), which is the average of the individual posterior probabilities of the subjects assigned to the group. An APP value greater than 70% for all groups suggests adequate classification[9]. Another criterion, the odds of correct classification (OCC), is the ratio of the odds of a correct classification in each group based on the maximum probability classification rule and on random assignment. A value greater than 5 for all groups indicates that the model has high assignment accuracy[9]. Finally, relative entropy is an overall measure of classification uncertainty, with values above 0.80 suggesting less classification uncertainty[10, 14].

**2.2 Bootstrap approach**

We now describe our proposed procedure to validate and quantify the uncertainty in the number of groups. We first use bootstrap to sample with replacement observations from the original data. Then, a LCGA (special case of LCGM) is fitted on each bootstrap sample and the number of groups is chosen independently in each sample using the BIC as described previously. This procedure thus allows checking if the number of groups found in the original data is also found across bootstrap samples. This procedure also allows quantifying if and how the number of groups varies from one bootstrap sample to another. In practice, we recommend that analysts report the bootstrap distribution of the number of groups and report the solutions found with each of the most frequent number of groups found using the bootstrap to reflect the estimated uncertainty.

**2.3 Simulation**

We evaluated the performance of our approach for different number and shapes of groups and for different sample sizes using simulated data. The purpose of the simulations is twofold: i) to verify whether the uncertainty identified by the bootstrap approach represents the sampling variability and ii) to determine whether the bootstrap approach can identify problems in model fitting that are not identified by other commonly used model adequacy criteria. To contrast the effect of different sample sizes and shapes on model performance, we considered scenarios with sample sizes of 100 and 500 subjects, 3 and 4 groups, curvilinear and intersecting groups. To reflect a typical application of LCGA based on a systematic literature, we considered the outcome as continuous with 5 follow-up times[15].

Datasets were generated for each group using a multivariate normal model. For all scenarios, the outcome Y is defined as follow:

$$Y_{it}^k = \beta_0^k + b_{0i}^k + \beta_1^k t + \beta_2^k t^2 + \cdots \beta_j^k t^j + \varepsilon_i^k$$

$$\varepsilon_i^k \sim N(0, \sigma_k^2)$$

$$b_{0i}^k \sim N(0, \sigma_{b_0^k}^2)$$

Due to the very high computational time, we limited the number of replications for each scenario to 100.

**2.4 Validation of identified groups**

For each bootstrap sample, the number of groups was selected using the BIC. Again, due to the computational time, we performed 100 bootstraps for each replication, which led to 10,000 datasets. To simplify the analyses, we focused only on the number of groups and considered a cubic polynomial shape for all groups. Next, we calculated the proportion of simulation replication samples where the correct number of groups was selected. This replication-wise variability represents the true sampling variability. Then, we computed the proportion of bootstrap samples where the correct number of groups was identified and compared the replication-wise variability to the bootstrap-estimated variability. We also compared the replication-wise variability to the Bayesian posterior probability of group numbers. The latter is an alternative criterion to select the number of groups that gives the probability that a model with *j* groups is the correct model among a set of models. Finally, we evaluated the ability of three commonly used model adequacy criteria (APP, OCC, and relative entropy) to identify uncertainty in the number of groups.

**2.5 Data analysis**

Simulations and analyses were performed using R.4.1.0 software (© 2021 The R Foundation for Statistical Computing, Vienna, Austria). Codes used for the simulation and the bootstrap approach are available in the Supplementary material. The trajectory groups were identified using the R package *FlexMix*[16]. This package implements a general framework for finite mixture models based on an Expectation-Maximization algorithm. For each scenario, we estimated models with number of groups varying from 1 to 5 and selected the optimal model. Then, we used the *boot* function for resampling. Finally, we computed the model adequacy criteria for each selected model from the replicated samples. Based on 100 replications, the maximal Monte Carlo standard error for estimating a proportion is 5%[17].

**3. RESULTS**

Figure 1 presents the simulated data for each scenario and for a replicated sample. Table 1 shows the results for all scenarios stratified by the population sample size. Among the 100 replicated samples, the correct number of groups was identified in at least 80% of cases, except in Scenario 2 (3 groups with 2 intersecting) with a sample size equals to 100. Overall, the proportion of bootstrap samples that identified the correct number of groups matched that of replicated samples; therefore, the bootstrap approach reproduced very well the true variability. The Bayesian posterior probability also performed well in 3 of the 4 scenarios. However, in scenario 2 with a sample size equals to 100, the average Bayesian posterior probability to identify the correct number of groups was 97% while only 71% of replicated samples identified the correct number of groups. Thus, it appears that the Bayesian posterior probability strongly underestimated the true variability. In terms of comparison with the commonly used model adequacy criteria, the bootstrap approach revealed uncertainty in the number of groups while all three model adequacy criteria were overall above their threshold, suggesting that the classification for each scenario was adequate, even when an incorrect number of groups was found (Table 2).

## 4. APPLICATION

Data from the Quebec Integrated Chronic Disease Surveillance System (QICDSS)[18] were used to identify patterns of medication claims between 2015 and 2018 in adults aged 75 to 80 years with diabetes. We chose to study this age group because older adults are prone to comorbidities[19, 20] and therefore to use many medications[21]. The QICDSS is a system built from data extracted from five medico-administrative files. The medication claim registry includes information on pharmaceutical services such as the name of the medications, quantity delivered and length of treatment. The death registry contains an indicator to identify the deceased and the cause of death. Health care information is provided by physician and hospital billing databases. The use of the data is approved by the *Commission d'accès à l'information du Québec*, the *Régie* *d'assurance de la maladie du Québec* and the *Ministère de la Santé et des Services sociaux* for surveillance purposes.

This illustrated example includes data on 46,235 participants aged 75 to 80 years who are permanently covered by the public drug insurance plan and have data for all four years. We considered models of 1 to 5 groups and selected the optimal model based on the BIC and the Bayesian posterior probability. Then, we computed model adequacy using the three commonly used criteria (APP, OCC, and relative entropy) and applied the bootstrap approach.

As shown in Table 3, the proportions of male and female were similar, and the median age was 77 years (Interquartile range (IQR)=76-79)). In 2015, individuals had a median number of comorbidities of 3 (IQR=2-6) and claimed a median annual number of different medications of 11 (IQR=8-16).

The BIC and the Bayesian posterior probability suggested the five-group model (Table S1). The three model adequacy criteria suggested an appropriate classification (Table S2). The identified groups were characterized by different levels of sustained use of medications over time (Figure 2). When applying our proposed bootstrap procedure with 100 bootstrap samples, the 5 groups solution was selected in each bootstrap sample (Table S2).

## 5. DISCUSSION

In this paper, we proposed a bootstrap approach to determine the uncertainty in the number of groups. For each generated scenario, our proposed approach reproduced the true variability very well while commonly used model adequacy criteria were unable to detect the uncertainty in the number of groups.

The bootstrap approach was also applied to real data and the number of groups (5 groups) identified was the same between the original data and the bootstrap samples. Therefore, the bootstrap approach allows us to represent the uncertainty of the number of groups in this real data. Using our proposed approach, coupled with commonly used model adequacy criteria, can increase confidence in the number of identified groups. Indeed, a recent study found that in some simulated scenarios, the commonly used model adequacy criteria failed to identify spurious groups and recommends the use of multiple criteria to better assess classification adequacy [22]. In addition, because the number of groups is influenced by several factors such as the sample size, number of time points and relative group size[7, 15], it is important to use different statistical approaches to assess group validity.

Based on our results, we suggest that the number of groups identified by the data should be considered and interpreted with caution. A concordance between the number of groups obtained with the bootstrap approach and those of the original dataset shows stability in the results. On the other hand, the bootstrap may reveal that there is substantial uncertainty in the true number of groups. In this case, we recommend that researchers present the results for different number of groups. As underlined by several authors, the use of cluster-based approach must build on existing theory to support the validity of groups[23, 24]. Several indicators such as pre-existing characteristics, subsequent outcomes, relationship of groups to outcomes or behaviors should differentiate the groups to assess their scientific utility[1].

Although from a descriptive perspective not recovering the correct number of groups may not be problematic, linking groups as exposure variables to health outcomes with the wrong number of groups may bias the results[1, 4]. Studies have shown that identifying groups that do not exist in the data can affect the determination of the effect of predictors by decreasing power and by obtaining a wrong predictor[1, 4].

The study has limitations. First, due to the computational challenges, we were limited in the number of scenarios to investigate. However, we contrasted the results for different number of groups and sample sizes. Second, the paper focused only on the selection of the number of groups. A future perspective is to adapt our approach to incorporate the selection of group shapes. Third, we considered only continuous outcomes, however, the extension to other distributions is straightforward. Other studies are needed to explore the performance of our proposed approach in other complex scenarios and other distributions. Finally, the number of groups was selected using the BIC because this is the most commonly used criterion in the literature. Other criteria such as the Lo-Mendall-Rubin Likelihood Ratio (LMR-LRT) and the parametric bootstrapped likelihood ratio test (BLRT) have been proposed to select the number of groups[10, 25], but they were not implemented in the *FlexMix* package in R.

In conclusion, our proposed bootstrap approach can be useful for studies on longitudinal finite mixture models as an internal validation approach to quantify the potential variability in the number of groups identified and especially for data with small sample sizes.

**Declaration of Conflicting Interests:** The authors declare that there is no conflict of interest.

**Funding:** This work was supported by a grant from the Natural Sciences and Engineering Research Council of Canada (NSERC) (grant number: RGPIN-2016-06295) and a Collaborative Health Research Projects grant from the Canadian Institute of Health Research and the NSERC (grant number: CPG-170621). Caroline Sirois holds a Junior 2 salary award from the Fonds de recherche du Québec – Santé (FRQS). Denis Talbot is supported by a Junior 1 salary award from the FRQS. Marc Sirois is supported by the FRQS, PhD fellowship.

|  **N=100**  |  **N=500**  |

Scenario 1: 3 curvilinear groups

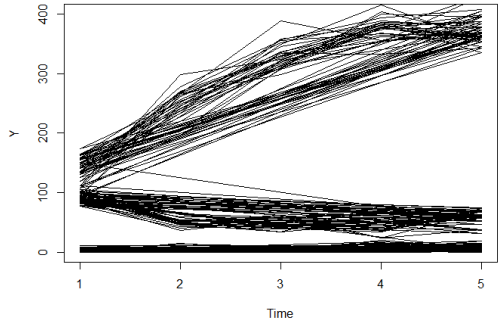 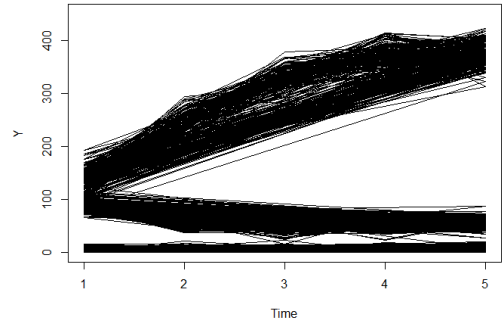

Scenario 2: 3 groups, 2 of which intersect

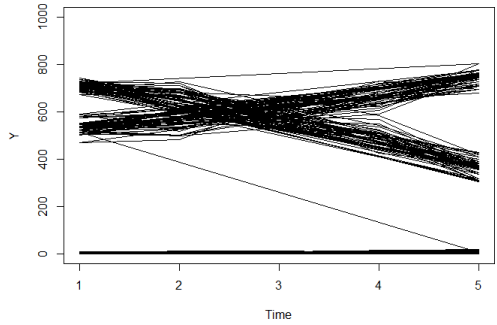 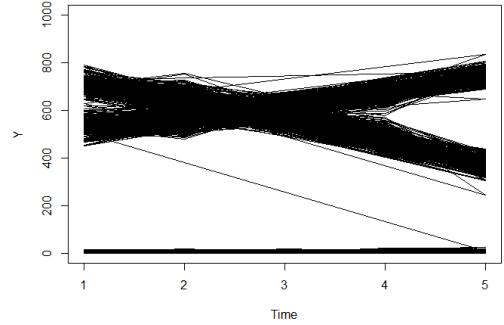

Scenario 3: 4 curvilinear groups

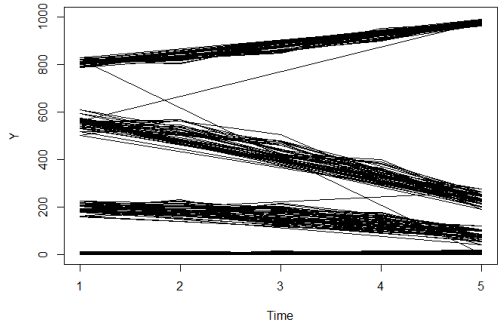 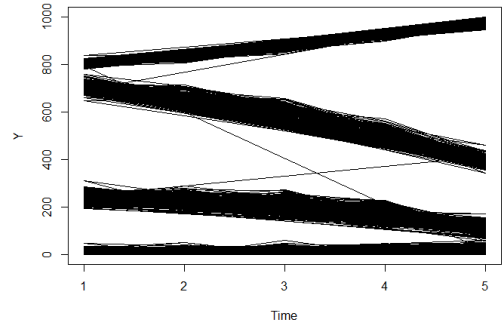

Scenario 4: 4 groups, 2 of which intersect

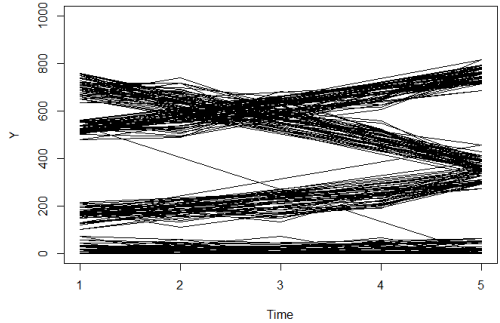 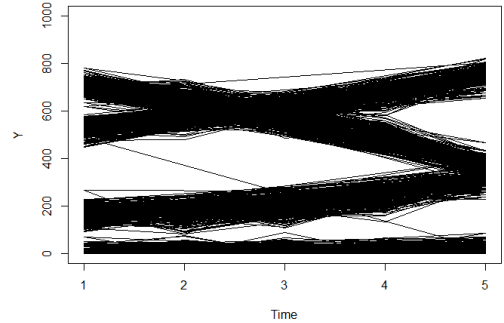

Figure 1: Simulated data for each scenario for one replicated dataset

Table 1: Fit indices for the trajectory model using simulated data by scenario

| Fit indices | Replications (100 samples) | |
|---|---|---|
| | Sample size=100 | Sample size=500 |
| **Scenario 1: 3 curvilinear groups** | | |
| *BIC* | | |
| Replication samples that identified the correct number of groups, % | 99.0 | 97.0 |
| *Bayesian posterior probability* | | |
| Probability of having identified the correct number of groups, mean | 97.0 | 94.9 |
| *Bootstrap* | | |
| Bootstrap samples that identified the correct number of groups, % | 93.6 | 95.9 |
| **Scenario 2: 3 groups, 2 of which intersect** | | |
| *BIC* | | |
| Replication samples that identified the correct number of groups, % | 71.0 | 94.0 |
| *Bayesian posterior probability* | | |
| Probability of having identified the correct number of groups, mean | 96.7 | 100.0 |
| *Bootstrap* | | |
| Bootstrap samples that identified the correct number of groups, % | 74.7 | 92.2 |
| **Scenario 3: 4 curvilinear groups** | | |
| *BIC* | | |
| Replication samples that identified the correct number of groups, % | 85.0 | 94.0 |
| *Bayesian posterior probability* | | |
| Probability of having identified the correct number of groups, mean | 87.3 | 92.4 |
| *Bootstrap* (100 samples) | | |
| Bootstrap samples that identified the correct number of groups, % | 86.1 | 93.9 |
| **Scenario 4: 4 groups, 2 of which intersect** | | |
| *BIC* | | |
| Replication samples that identified the correct number of groups, % | 92.0 | 80.0 |
| *Bayesian posterior probability* | | |
| Probability of having identified the correct number of groups, mean | 92.0 | 86.7 |
| *Bootstrap* (100 samples) | | |
| Bootstrap samples that identified the correct number of groups, % | 91.4 | 90.0 |

*Legend.* For each scenario, the percentage of the three indices (BIC, Bayesian posterior probability and the bootstrap) is expected to be similar. The similarity of the Bayesian posterior probability and bootstrap to the BIC means that both criteria were able to detect the uncertainty in the number of groups.

Table 2: Model adequacy criteria for each scenario

| Criteria | Threshold | Scenario 1: 3 curvilinear groups | |
| --- | --- | --- | --- |
| | | Sample size=100 | Sample size=500 |
| *Average posterior probability* | | | |
| Replication samples above threshold, n (%) | ≥ 70% | 100 (100) | 99 (99) |
| Estimated values, minimum - maximum | | 1.0 – 1.0 | 0.6 – 1 |
| *Odds of correct classification* | | | |
| Replication samples above threshold, n (%) | ≥ 5 | 100 (100) | 100 (100) |
| Estimated values, minimum - maximum | | Inf. – Inf. | 17 – Inf. |
| *Relative entropy* | | | |
| Replication samples above threshold, n (%) | ≥ 80% | 100 (100 | 100 (100) |
| Estimated values, minimum - maximum | | 1.0 – 1.0 | 0.9 – 1.0 |
| | | Scenario 2: 3 groups, 2 of which intersect | |
| | | Sample size=100 | Sample size=500 |
| *Average posterior probability* | | | |
| Replication samples above threshold, n (%) | ≥ 70% | 100 (100) | 100 (100) |
| Estimated values, minimum - maximum | | 0.9 – 1.0 | 0.8 – 1.0 |
| *Odds of correct classification* | | | |
| Replication samples above threshold, n (%) | ≥ 5 | 100 (100) | 100 (100) |
| Estimated values, minimum - maximum | | 30.0 – Inf. | 35.0 – Inf. |
| *Relative entropy* | | | |
| Replication samples above threshold, n (%) | ≥ 80% | 100 (100) | 100 (100) |
| Estimated values, minimum - maximum | | 0.9 – 1.0 | 0.9 – 1.0 |
| | | Scenario 3: 4 curvilinear groups | |
| | | Sample size=100 | Sample size=500 |
| *Average posterior probability* | | | |
| Replication samples above threshold, n (%) | ≥ 70% | 100 (100) | 100 (100) |
| Estimated values, minimum - maximum | | 1.0 – 1.0 | 0.7 – 1.0 |
| *Odds of correct classification* | | | |
| Replication samples above threshold, n (%) | ≥ 5 | 100 (100) | 100 (100) |
| Estimated values, minimum - maximum | | 30.0 – Inf. | 15.8 – Inf. |
| *Relative entropy* | | | |
| Replication samples above threshold, n (%) | ≥ 80% | 100 (100) | 100 (100) |
| Estimated values, minimum - maximum | | 1.0 – 1.0 | 0.91 – 1.0 |
| | | Scenario 4: 4 groups, 2 of which intersect | |
| | | Sample size=100 | Sample size=500 |
| *Average posterior probability* | | | |
| Replication samples above threshold, n (%) | ≥ 70% | 100 (100) | 100 (100) |
| Estimated values, minimum - maximum | | 0.9 – 1.0 | 0.7 – 1.0 |
| *Odds of correct classification* | | | |
| Replication samples above threshold, n (%) | ≥ 5 | 100 (100) | 100 (100) |
| Estimated values, minimum - maximum | | 43.0 – Inf. | 17.3– Inf. |
| ***Relative entropy*** | | | |
| Replication samples above threshold, n (%) | ≥ 80% | 100 (100) | 100 (100) |
| Estimated values, minimum - maximum | | 1.0 – 1 | 0.9 – 1 |

***Abbreviations.*** Inf. : Infinity

Table 3: Characteristics of included individuals with diabetes aged between 75-80 years in 2015

| | Included individuals (N = 46,235) |
|---|---|
| Male, n (%) | 23,356 (50.5) |
| Age, median (IQR) | 77.28 (76.1 – 78.6) |
| Residence area, n (%) | |
|     Urban | 36,576 (79.1) |
|     Rural | 8181 (20.8) |
| Material deprivation index, n (%) | |
|     Quintile 1 (most privileged) | 6221 (15.0) |
|     Quintile 2 | 6993 (16.8) |
|     Quintile 3 | 8418 (20.2) |
|     Quintile 4 | 9493 (22.8) |
|     Quintile 5 (most deprived) | 10,487 (25.2) |
| Social deprivation index, n (%) | |
|     Quintile 1 (most privileged) | 7183 (17.3) |
|     Quintile 2 | 8031 (19.3) |
|     Quintile 3 | 8368 (20.1) |
|     Quintile 4 | 8948 (21.5) |
|     Quintile 5 (most deprived) | 9082 (21.8) |
| Number of medications claimed in the preceding year, median (IQR) | 11 (8.0 – 16.0) |
| Number of comorbidities, median (IQR) | 3 (2.0 – 6.0) |

***Abbreviations***. IQR : Interquartile range

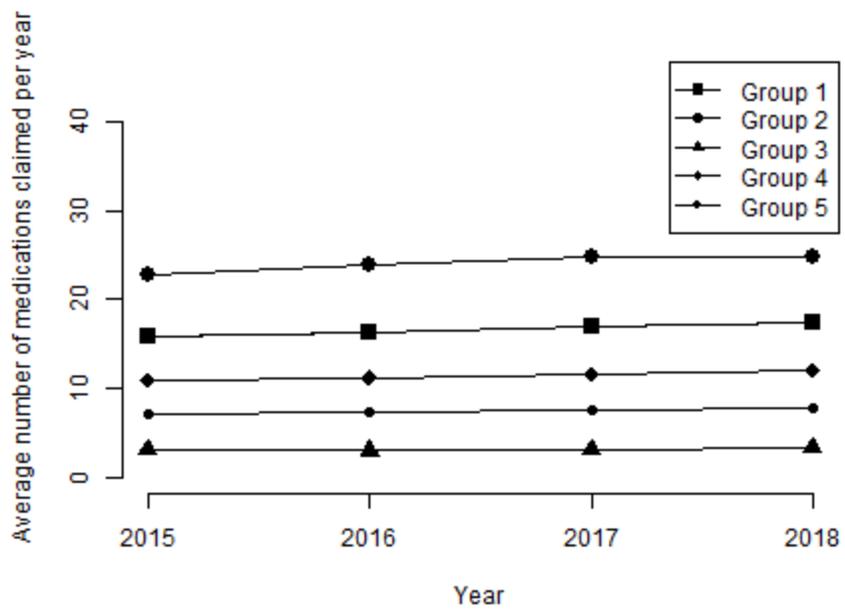

Figure 2: Trajectories of prescribed medications among individuals with diabetes aged between 75-80, 2015-2018

**Supplementary material**

Table S1: Fit indices of models from 1 to 5 groups (original data), Quebec Integrated Chronic Disease Surveillance System (QICDSS), n=46,235 participants

| Models | BIC | Bayesian posterior probability |
| --- | --- | --- |
| 1 group | 1,206,553 | 0 |
| 2 groups | 1,109,320 | 0 |
| 3 groups | 1,063,390 | 0 |
| 4 groups | 1,040,358 | 0 |
| 5 groups | 1,027,156 | 1 |

Table S2: Model adequacy criteria for the trajectory model for medications use, Quebec Integrated Chronic Disease Surveillance System (QICDSS), n=46,235 participants

| | | Groups | | | | |
| --- | --- | --- | --- | --- | --- | --- |
| | | 1 (n= 10,621) | 2 (n= 12,243) | 3 (n= 15,682) | 4 (n= 4507) | 5 (n= 3182) |
| *Criteria* | | | | | | |
| Average posterior probability | | 0.9 | 0.9 | 0.9 | 0.9 | 0.9 |
| Odds of correct classification | | 30.3 | 23.7 | 16.7 | 106.9 | 192.1 |
| Relative entropy | 0.8 | | | | | |
| *Bootstrap approach* | | | | | | |
| Samples that identified 5 groups, % | 100 | | | | | |